# Automatic Event Recognition Employing Optically Excited Electro-Nuclear Spin Coherence

SHAMIMA A. MITU,[1] XI SHEN,[1] JULIAN GAMBOA,[1] TABASSOM HAMIDFAR,[1]
AND SELIM M. SHAHRIAR[1,2]

[1]*Department of Electrical and Computer Engineering, Northwestern University, Evanston, IL 60208, USA*
[2]*Department of Physics and Astronomy, Northwestern University, Evanston, IL 60208, USA*
*\* shahriar@northwestern.edu*

**Abstract:** An analog automatic event recognition (AER) system can be realized by combining the technique of holographic image recognition with the process of temporal signal correlation employing stimulated photon echo in an ensemble of two-level atoms. For efficient operation of the AER system, the optical transition in the two-level system must have a large oscillator strength. However, this implies a rapid decay of the excited state, which highly constrains the duration of the event that can be recognized, even when the images are retrieved rapidly from a fast parallel optical memory device. This constraint can be overcome by using a three-level Lambda system to transfer optical coherence to long-lived electro-nuclear spin coherence and vice-versa. Here, we present an analysis of the AER process employing such a Lambda system. We also show that due to the longitudinal and transverse variation of the phases of the optical fields undergoing Fresnel diffraction, the density matrix equations must account for the differences in the phases of the coherence excited in each pixel and the corresponding phases of the optical fields in subsequent pulses.

## 1. Introduction

Non-linear interactions induced by multi-photon transitions in atomic media can be exploited to enhance significantly the capabilities of many imaging and image processing systems. This is due to the convergence of several factors. First, the use of resonant or nearly-resonant excitation fields makes the atomic non-linearity very strong. Second, the energy levels in atoms are very well understood, and follow stringent selection rules, thus making it possible to tailor the non-linear response for specific applications. Third, atoms can store temporal information efficiently in the form of long-lived electro-nuclear spin coherence induced by optical fields. Previously, we propose the use of such interactions for an application of significant interest: ultrafast Automatic Event Recognition (AER) [1,2]. Specifically, the AER system can be realized by the technique of Spatio-Temporal Holographic Correlation (STHC) [1,2,3,4,5]. The basic physical processes that need to be combined to realize the AER system have been studied earlier [6,7,8,9,10].

Recent advances in communication, computing and surveillance, in both civilian and military sectors, have led to a massive growth in video content. A key challenge in this context is the need to identify an event, embodied in a short video clip (VC), in a much larger video data base.  Conventional techniques [11,12] using digital computers require a very time consuming pre-processing of the data in order to perform this task.  A similar challenge also exists in the field of recognizing a single image in a large collection of images.  In the latter case, again, conventional techniques require significant pre-processing of the data that takes a long time.  It is well known that the spatial parallelism of optics can perform image recognition very fast and efficiently, using a holographic correlator, without requiring any pre-processing [13,14,15,16,17,18,19,20,21,22].

To see what is needed to extend this approach to the task of event recognition, we recall first the process of holographic image recognition.  Briefly, the query image is Fourier Transformed (FT'd) with a lens, and its interference with a plane wave is recorded, in the form of spatial gratings, in a holographic substrate.  The images from the data base are also FT'd using a lens, and sent to the holographic substrate to produce multiplication.  Inverse FT of the resulting signal yields a correlation peak in case of a match.  Because of the spatial invariance of the FT process, this system works in a translation invariant manner.  Scale and rotation invariance can also be achieved if Polar Mellin Transform (PMT) is added to the process [18,20,21,22].

A similar approach can be used for finding a query signal in a one dimensional data stream (such as recognizing a word in a segment of speech).  Interference between a reference signal and the query signal are stored in a spectrally broad-band dynamic memory. Because of the spectral spread in the elements of this memory, what gets recorded is automatically the FT of the query signal.  When the data stream is sent to the memory, a non-linear interaction produces a product between the FT of the query signal and that of the data stream.  The resulting temporal signal produced by the medium automatically produces the inverse FT of this product, yielding a correlation peak in case of a match [1].  This process is also translation invariant, and can be rendered scale invariant in addition, by using one dimensional Mellin Transform of the signals as inputs [23].

Consider now the case of a video clip (VC).  It is a function of two spatial dimensions and one temporal dimension.  Thus, in order to recognize it, what is needed is a medium that can record the spectral FT, in addition to the spatial FT.  Such a medium would require the following spatio-spectral properties.  First, the active elements in the medium must have a spread in the spectral response, in order to record both the amplitude and the phase of the spectral FT components with high fidelity. Since both the amplitude and the phase information must be retained over the duration of the time for performing the VC recognition task, the recording must be done in the form of induced oscillating dipole antennae. Second, the active elements in the medium must stay confined to a spatial area no larger than the spatial resolution of the FT of the images in the VC during the time for performing the VC recognition task.

The maximum duration of the event that can be recognized using the AER system is limited by the lifetime of the coherence created in the atomic ensemble. As such, we had proposed [1] a system where the query event is stored in an optical memory, such as a crystal of lithium niobate, using holographic techniques, and retrieved rapidly via spatial and angular multiplexing with an acousto-optic modulator. In addition, it is necessary to use an optical transition that has an excited state with a very long lifetime, such as those used for optical clocks. However, since the decay rate of an excited state is proportional to the square of the dipole matrix element (DME) for coupling this to the ground state, it follows that the DME for these transitions are very small. This in turn implies that extremely intense lasers would be needed to produce the maximum dipole moment, and the signal radiated by these dipoles would be very small.

To overcome this constraint, we had proposed [1,2] the use of a $\Lambda$ type three level system, where the two ground states would, for example, be long-lived hyperfine states within the ground electronic state of an alkali atom, such as $^{87}$Rb. Specifically, an optical field is always present on one leg of the $\Lambda$ system, and the signal optical fields are applied along the other leg, ensuring that the frequency difference between the two optical fields match the hyperfine splitting. Under such a condition, the optical coherence on the signal leg can be transferred to electo-nuclear spin coherence among the hyperfine states, and vice versa. In the numerical simulations reported in our previous work [], we considered only the case of an artificially long-lived two level system, in order to elucidate the basic concept. In this paper, we consider the realistic case where a $\Lambda$ type three level system is used for simulating the expected behavior of the AER system. We show how the fidelity of the AER process depends on various operating parameters, including the optical Rabi frequencies, the optical detunings, and the decay rate of the intermediate state. We also show that due to the longitudinal and transverse variation of the phases of the optical fields undergoing

Fresnel diffraction, the density matrix equations must account for the differences in the phases of the coherence excited in each pixel and the corresponding phases of the optical fields in subsequent pulses.

**2. Translation Invariant Temporal Correlator**

To illustrate the temporal correlator, consider a medium with inhomogeneous broadening, where the atoms have a range of resonant frequencies. In such a medium, a temporal data sequence can be recorded. The process begins with the application of a write pulse, followed by the query data stream with a certain time delay. The spectral domain interference between the writing beam and the query data stream is encoded in the coherence produced within the atomic medium. When the reference data stream is introduced into the system, a correlation peak is observed in a temporally shift-invariant manner. In what follows, we consider three different atomic systems: (1) An idealized, decay-free optical two-level atomic system, (2) A three-level system incorporating decays, and (3) an effective two-level spin transition system resulting from adiabatic elimination of the excited state in the three level system.

*2.1 An Optical Two Level System without Decay*

Consider an ensemble of two-level atoms excited by a laser field with frequency ω. The energy levels, |1> and |2>, are coupled by a laser field characterized by a Rabi frequency $\Omega_0$ and a detuning δ. Under the rotating wave approximation and transformation, and ignoring the effect of decay of the excited state, the Schrödinger equation can be expressed as follows:

$$i\hbar \frac{\partial |\Psi\rangle}{\partial t} = H|\Psi\rangle; \quad H = \frac{\hbar}{2}\begin{bmatrix} 0 & \Omega_0 \\ \Omega_0 & -2\delta \end{bmatrix}, \tag{1}$$

where $|\psi\rangle = c_1|1\rangle + c_2|2\rangle$ and $\delta = \omega - (E_e - E_g)/\hbar$.

The general solution to this equation is given by:

$$\begin{bmatrix} \tilde{c}_1(t+\tau) \\ \tilde{c}_2(t+\tau) \end{bmatrix} = e^{\frac{i\delta\tau}{2}} \begin{bmatrix} \cos(\theta) - i\delta'\sin(\theta) & -i\Omega_0/\Omega'\sin(\theta) \\ -i\Omega_0/\Omega'\sin(\theta) & \cos(\theta) + i\delta'\sin(\theta) \end{bmatrix} \begin{bmatrix} \tilde{c}_1(t) \\ \tilde{c}_2(t) \end{bmatrix}, \tag{2}$$

where $\theta = \Omega'\tau/2$ and $\Omega' = \sqrt{\Omega_0^2 + \delta^2}$, with $\Omega_0$ being the Rabi frequency and $\delta$ being the detuning.

The atomic spectral distribution is assumed to have a Gaussian profile with a width of $\Delta\omega$, resulting from Doppler broadening. For our simulations, we assume the laser is resonant with stationary atoms, meaning $\delta=0$ and $\delta_{eff}=-kv$. We suppose that initially all N atoms are prepared in state 1>. We first determine the quantum state of a band of atoms with a velocity centered at some value v, after it has interacted with several laser pulses in sequence. This state is then used to determine the amplitude and phase of the induced dipole moment (proportional to $\rho_{12}$) for this band, and add them together, weighted by the Gaussian distribution. The resulting net dipole moment induced by the pulses is given by: $P(t_4) = \int_{-\infty}^{\infty} \rho_{12}(t_4,\delta) g(\delta) d\delta$, where $g(\delta)$ is the Gaussian spectral distribution.

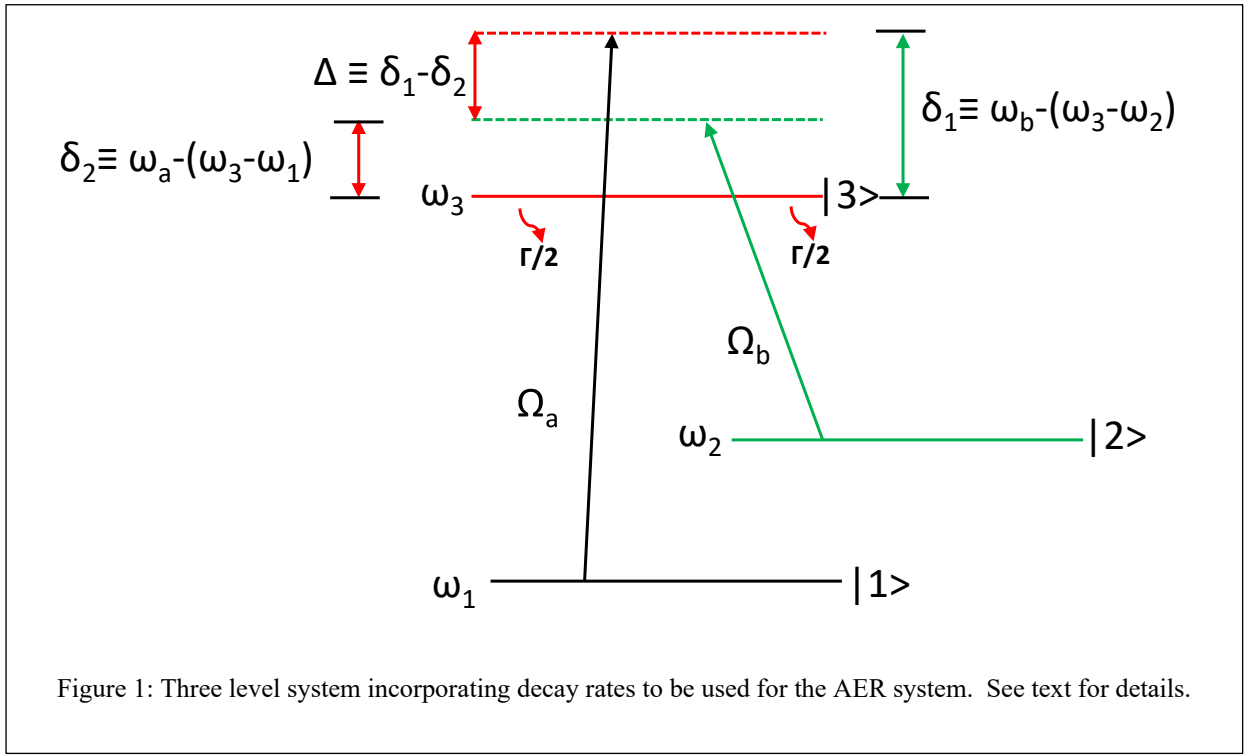

Figure 1: Three level system incorporating decay rates to be used for the AER system. See text for details.

*2.2 Three level system*

Consider next a three level system as the atomic medium, as illustrated in Figure 1. Level |1> and level |3> are coupled by Rabi frequency $\Omega_a$. Level |2> and |3> are coupled by Rabi frequency $\Omega_b$. For all the processes to be investigated using such a system, it is assumed that the field that couples levels |2> and |3> is always present. Under the rotating wave approximation and transformation, the Schrödinger equation can be expressed as follows [24]:

$$i\hbar \frac{\partial |\Psi\rangle}{\partial t} = H|\Psi\rangle; \quad H = \frac{\hbar}{2}\begin{bmatrix} \Delta & 0 & \Omega_a \\ 0 & -\Delta & \Omega_b \\ \Omega_a & \Omega_b & -2\delta - i\Gamma \end{bmatrix}, \quad (3)$$

where $|\Psi\rangle = c_1|1\rangle + c_2|2\rangle + c_3|3\rangle$, $\Delta = \delta_1 - \delta_2$, and $\delta = (\delta_1 + \delta_2)/2$. This equation can be used to model the transient behavior of the system if the common mode detuning, $\delta$, is large enough so that the population of state |3> remains very small. For a more general situation, it is necessary to take into account the fact that level |3> will decay to both level |1> and |2> at the rate of $\Gamma/2$. This can be done by using the density matrix equations of motion, which can be expressed as [24]:

$$\begin{bmatrix} \dot{\tilde{\rho}}_{11} \\ \dot{\tilde{\rho}}_{12} \\ \dot{\tilde{\rho}}_{13} \\ \dot{\tilde{\rho}}_{21} \\ \dot{\tilde{\rho}}_{22} \\ \dot{\tilde{\rho}}_{23} \\ \dot{\tilde{\rho}}_{31} \\ \dot{\tilde{\rho}}_{32} \\ \dot{\tilde{\rho}}_{33} \end{bmatrix} = \frac{1}{2}\begin{bmatrix} 0 & 0 & i\Omega_a & 0 & 0 & 0 & -i\Omega_a & 0 & \Gamma \\ 0 & -i2\Delta & i\Omega_b & 0 & 0 & 0 & 0 & -i\Omega_a & 0 \\ i\Omega_a & i\Omega_b & -i2\delta_1 - \Gamma & 0 & 0 & 0 & 0 & 0 & -i\Omega_a \\ 0 & 0 & 0 & -i2\Delta & 0 & i\Omega_a & -i\Omega_b & 0 & 0 \\ 0 & 0 & 0 & 0 & 0 & i\Omega_b & 0 & -i\Omega_b & \Gamma \\ 0 & 0 & 0 & i\Omega_a & i\Omega_b & -i2\delta_2 - \Gamma & 0 & 0 & -i\Omega_b \\ -i\Omega_a & 0 & 0 & -i\Omega_b & 0 & 0 & i2\delta_1 - \Gamma & 0 & i\Omega_a \\ 0 & i\Omega_a & 0 & 0 & -i\Omega_b & 0 & 0 & i2\delta_2 - \Gamma & i\Omega_b \\ 0 & 0 & -i\Omega_a & 0 & 0 & -i\Omega_b & i\Omega_a & i\Omega_b & -2\Gamma \end{bmatrix}\begin{bmatrix} \tilde{\rho}_{11} \\ \tilde{\rho}_{12} \\ \tilde{\rho}_{13} \\ \tilde{\rho}_{21} \\ \tilde{\rho}_{22} \\ \tilde{\rho}_{23} \\ \tilde{\rho}_{31} \\ \tilde{\rho}_{32} \\ \tilde{\rho}_{33} \end{bmatrix} \quad (4)$$

For this three level system, the resulting net dipole moment along the 1-3 transition induced by the pulse sequence is given by: $P(t_4) = \int_{-\infty}^{\infty} \rho_{13}(t_4, \delta) g(\delta) d\delta$, where g($\delta$) is the Gaussian spectral distribution.

## 2.3 Effective two level system

The three level system can be reduced to an effective two-level system when the detuning δ is sufficiently large [25]. The effective two-level system is produced via adiabatic elimination of state |3>, starting with Eqn. (3), and then adding source terms to the populations of states |1> and |2> in order to conserve the number of atoms in the system. The resulting density matrix equations for the effective two-level system are as follows:

$$\begin{bmatrix} \dot{\tilde{\rho}}_{11} \\ \dot{\tilde{\rho}}_{12} \\ \dot{\tilde{\rho}}_{21} \\ \dot{\tilde{\rho}}_{22} \end{bmatrix} = \frac{1}{2} \begin{bmatrix} -2\Gamma_{eff} & i\Omega_{eff} & -i\Omega_{eff} & 2\Gamma_{eff} \\ i\Omega_{eff} & -2\Gamma_{eff} - i2\Delta & 0 & -i\Omega_{eff} \\ -i\Omega_{eff} & 0 & -2\Gamma_{eff} + i2\Delta & i\Omega_{eff} \\ 2\Gamma_{eff} & -i\Omega_{eff} & i\Omega_{eff} & -2\Gamma_{eff} \end{bmatrix} \begin{bmatrix} \tilde{\rho}_{11} \\ \tilde{\rho}_{12} \\ \tilde{\rho}_{21} \\ \tilde{\rho}_{22} \end{bmatrix}, \quad (5)$$

Here $\Omega_{eff} = \Omega_a^2/(2\delta + i\Gamma)$ and $\Gamma_{eff} = \Omega_a^2 \Gamma/(4\delta^2 + \Gamma^2)$.

## 2.4 Results

We now consider the process of shift-invariant temporal signal recognition. In works in the manner illustrated in Figure 2. Briefly, we start with applying the recording pulse, at time T1, which is denoted as a(t) in the top panel of Figure 2. This is followed by application of another pulse, which represents the query signal, denoted as b(t), at time T2. At a later time, T3, we apply the reference pulse, denoted as c(t). Each pulse is Gaussian in shape, and is chosen to have an area, defined as the integral of the product of the Rabi frequency and time over the duration of the pulse, of $\pi/5$. The goal is to generate an output signal, at a time T4 such that (T4-T3)=(T2-T1), if the query and reference pulses match in their temporal shapes. Here, we consider the process in its simplest form, where the query and reference signals are identical pulses. As such, it corresponds to the process of an ideal stimulated photon echo. In the bottom panel of Figure 2., we show the results of simulation of this process using the ideal two level system without decay, which makes use of Eqn. (1). As can be seen, there is a clear echo signal generated at time T4= T3+T2-T1, which satisfies the condition stated above. Note that this process is translation invariant, meaning that the echo will appear at the proper time independent of the time separation of (T2-T1), satisfying the condition that (T4-T3)=(T2-T1). This translation invariance is a manifestation of the fact that the spectra of the Fourier Transforms of b(t) and c(t) are identical, except for phase shifts.

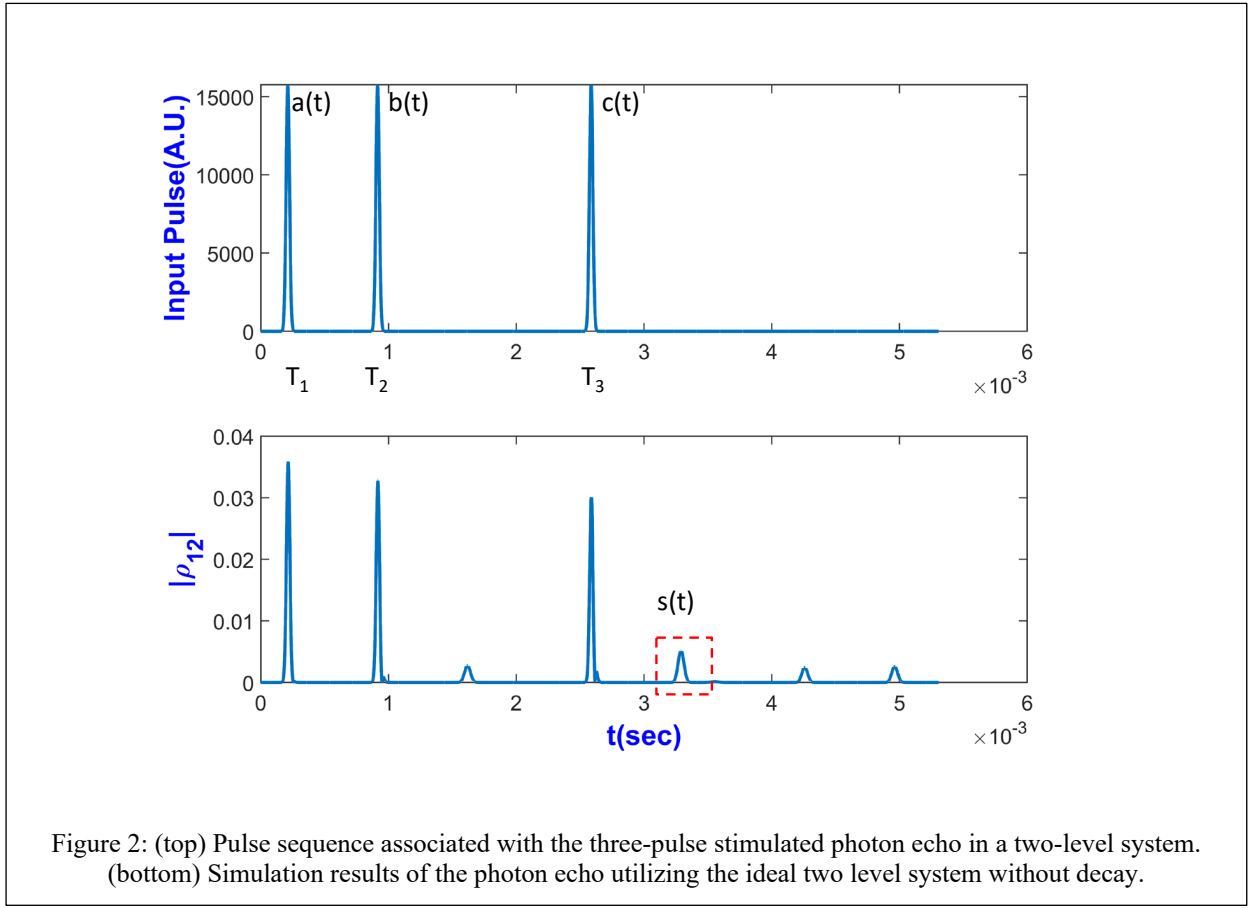

Figure 2: (top) Pulse sequence associated with the three-pulse stimulated photon echo in a two-level system. (bottom) Simulation results of the photon echo utilizing the ideal two level system without decay.

Of course, there are some additional, smaller signals that also appear, which can be understood as direct photon echoes, rather than stimulated photon echoes. The first one, appearing at time T2+(T2-T1) is the echo of a(t) produced by b(t). The second one, occurring at T3+(T3-T2) is the echo of b(t) produced by c(t). The third one, occurring at T3+(T3-T1) is the echo of a(t) produced by c(t). In the automatic event recognition system, the window for detecting the desired correlation signal will be constrained to exclude these additional echo signals [1,2].

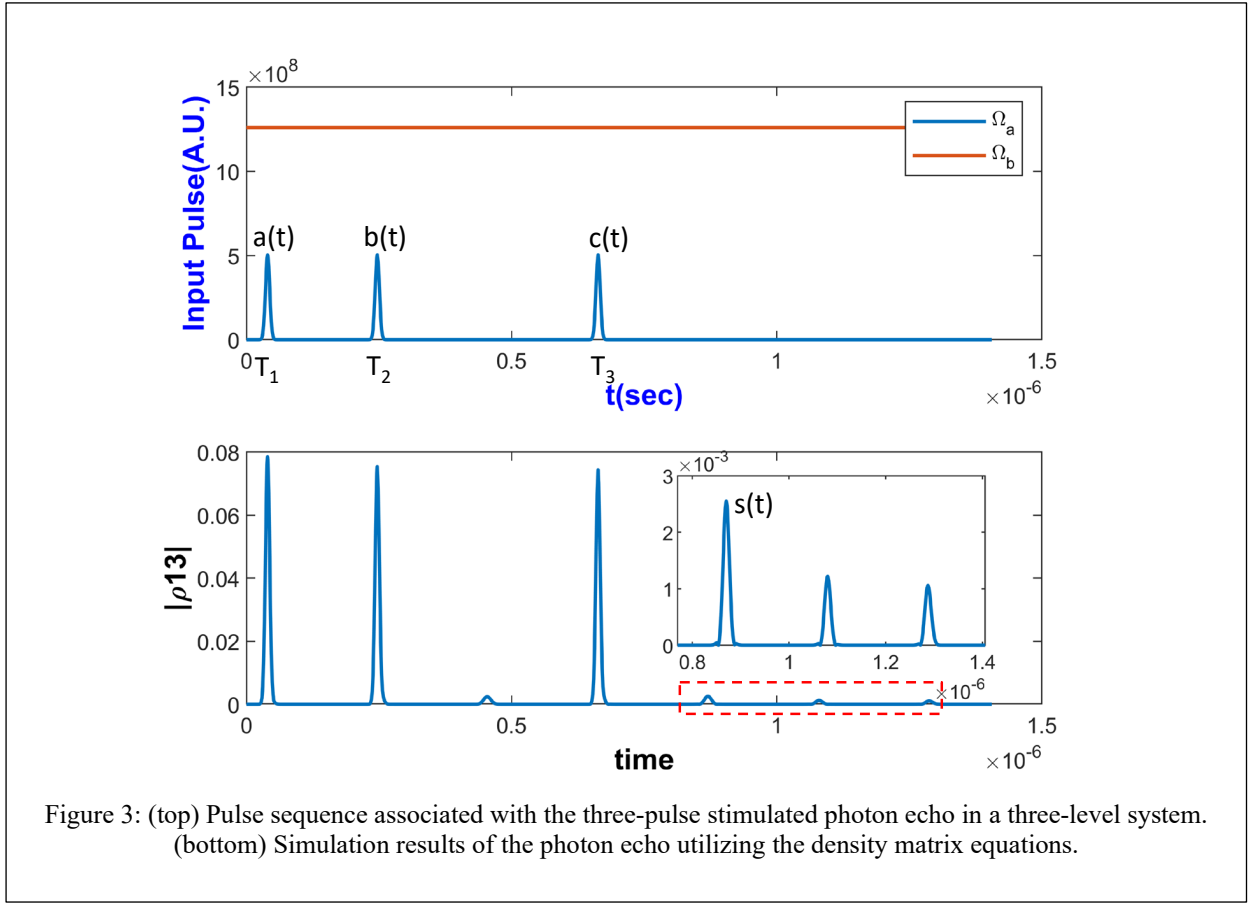

Figure 3: (top) Pulse sequence associated with the three-pulse stimulated photon echo in a three-level system. (bottom) Simulation results of the photon echo utilizing the density matrix equations.

Next, we consider the same process, but use the three-level system to simulate the response. The pulse sequence is illustrated in the top panel of Figure 3. Here, the top panel displays the input signals (write pulse, query pulse and reference pulse), applied along the 1-3 transition. The field along the 2-3 transition is assumed to be present at all times. $\Omega_a$ is the Rabi frequency of coupling rate between level |1> and |3>, and $\Omega_b$ is the Rabi frequency of coupling rate between level |2> and |3>. Each pulse is Gaussian in shape, and has an area of $\pi/10$. The bottom panel of Figure 3 shows the results of the simulation, generated using the density matrix model for the three level system, as shown in Eqn. (4). Here, the signal is expressed in terms of the amplitude of the dipoles excited on the 1-3 transition: $|\rho_{13}|$. The simulated photon echo signal appears at time $T_3+T_2-T_1$ as expected. Just as in the case of the two-level model, we see the additional echo signals as well. It should be noted that the amplitude of $|\rho_{13}|$ is significantly smaller than the signal shown earlier for the case of an ideal two-level system. The reason for this is as follows. The actual echo in this case produces a strong signal for $|\rho_{12}|$, as we will show next, using the effective

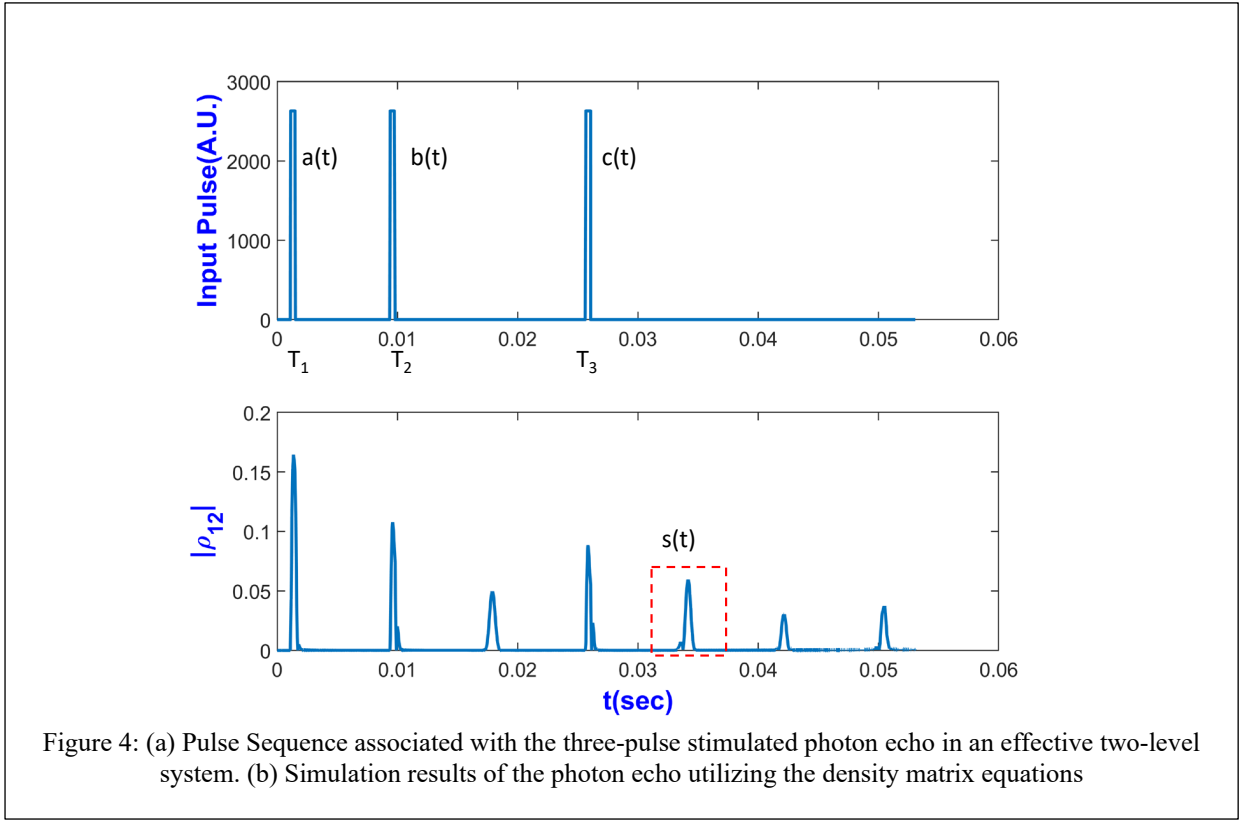

Figure 4: (a) Pulse Sequence associated with the three-pulse stimulated photon echo in an effective two-level system. (b) Simulation results of the photon echo utilizing the density matrix equations

two-level model. The magnitude of $|\rho_{13}|$ is proportional to $|\rho_{12}|$, but with a proportionality constant that depends on the Rabi frequency on the 2-3 transition, and the value of common mode detuning. However, since the actual signal we want to observe has to be in the optical domain, we must use $|\rho_{13}|$, and not $|\rho_{12}|$, as the signal. This is a necessary price to pay for using the three level system.

We now consider the same process again, but carry out the simulation using the effective two level model derived from the three level system. This is illustrated in Figure 4. Here, each of the three pulses (write, query, reference) is a rectangular one, with an area of $\pi/3$, which is defined as the product of the effective Rabi frequency and the pulse duration. The simulation is carried out using the density matrix version of the effective two level system, as shown in Eqn. (5) . The results are shown in the bottom panel of Figure 4. As can be seen, the primary signal corresponding to the stimulated photon echo appears at time T3+T2-T1, as expected. We also see the additional echo signals appearing at the expected times. Note that here the signal is represent by the magnitude of the coherence along the 1-2 transition: $|\rho_{12}|$. As noted above, this is much bigger than $|\rho_{13}|$. However, since the 1-3 transition would emit microwave radiation, it is not suitable for detecting optical correlation for the AER system. As such, we have to consider the magnitude of $|\rho_{13}|$ as the signal of interest, shown earlier in the bottom panel of Figure 3.

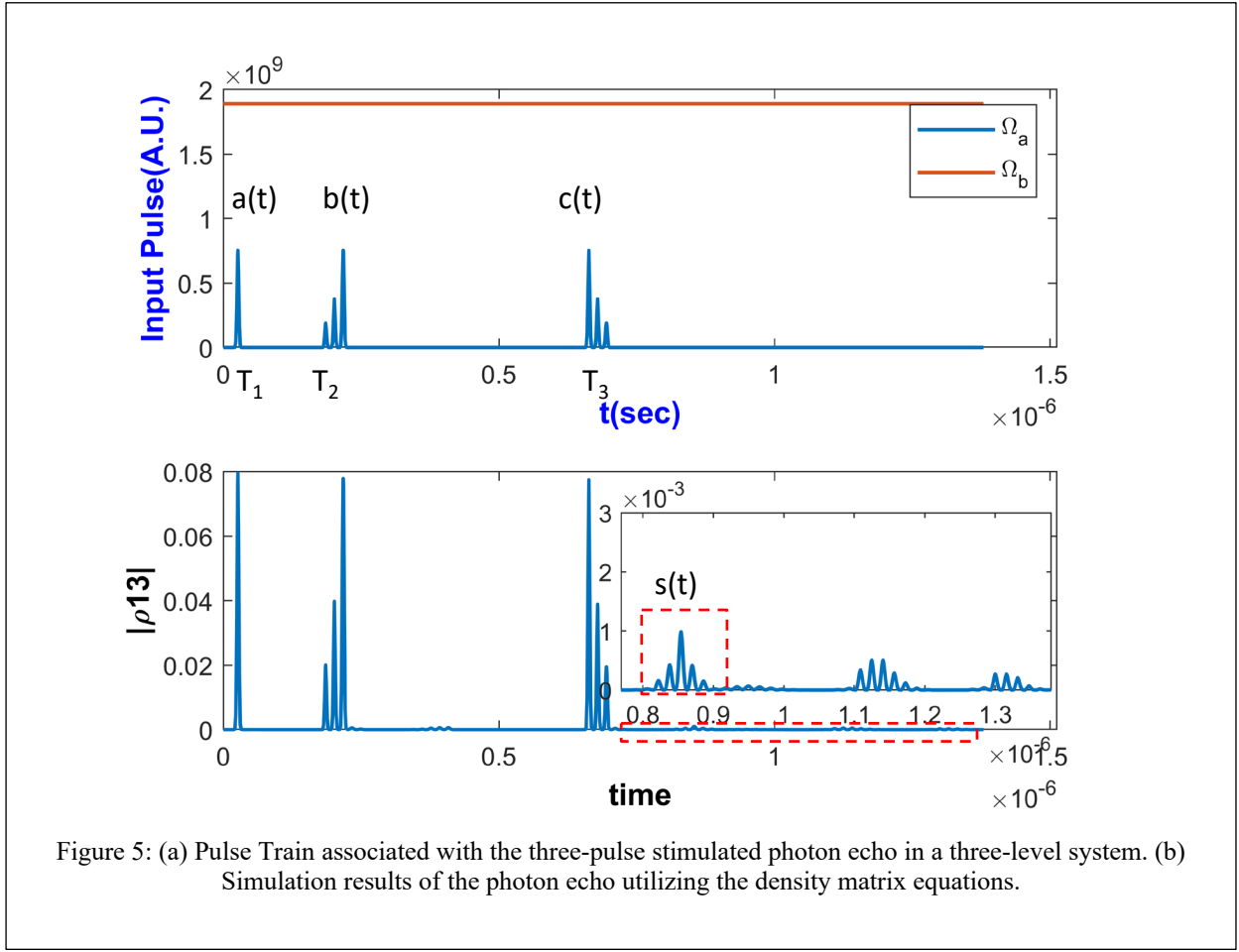

Figure 5: (a) Pulse Train associated with the three-pulse stimulated photon echo in a three-level system. (b) Simulation results of the photon echo utilizing the density matrix equations.

Finally, we repeat the process by replacing the simple query and reference pulses with packets of pulses, as illustrated in the top panel of Figure 5. For technical reasons [2], the sequence within the reference packet has to be in reverse order of the corresponding sequence in the query packet for maximum correlation. The simulation is carried out with the density matrix equations for the three level system, as shown in Eqn. (4). The results are shown in the bottom panel of Figure 5. As can be seen, the primary signal corresponding to the correlation appears at time T3+T2-T1, as expected. We also see the additional echo signals appearing at the expected times.

## 3. Automatic Event Recognition System

The basic configuration for the AER system is illustrated schematically in Figure 6. The input signal consists of three key components: a writing pulse, a query signal (a video segment) and a reference signal (another video segment). First, the writing pulse is applied to the system which propagates through the first lens, with a focal length L, and then reaches the atomic medium plane at time T1. After a certain time lag, the query signal propagates through the first lens and reaches the atomic plane at time T2. Following another delay, the reference signal is sent, which also propagates through the first lens, and reach the atomic medium plane at time T3. For each of these signals, the first lens produces a spatial Fourier Transform (FT) of the signal at the SLM at the plane of the atomic medium. At time T3+T2-T1, the correlation signal appears in the atomic medium plane. This signal propagates through the second lens, also with a focal length of L, and the three-dimensional correlation result between the query signal and the reference signal appears in the detector plane. Note that the second lens also produces a spatial FT of the signal at the output of the atomic medium at the detector plane. The writing pulse is a small circular region in the SLM so that the spatial FT of the pulse is a plane wave in the atomic medium. Additionally, the duration of this writing pulse needs to be kept short enough to ensure the temporal Fourier spectrum is wider than those of the query and reference signals. Additional details of this process can be found in refs. [1] and [2].

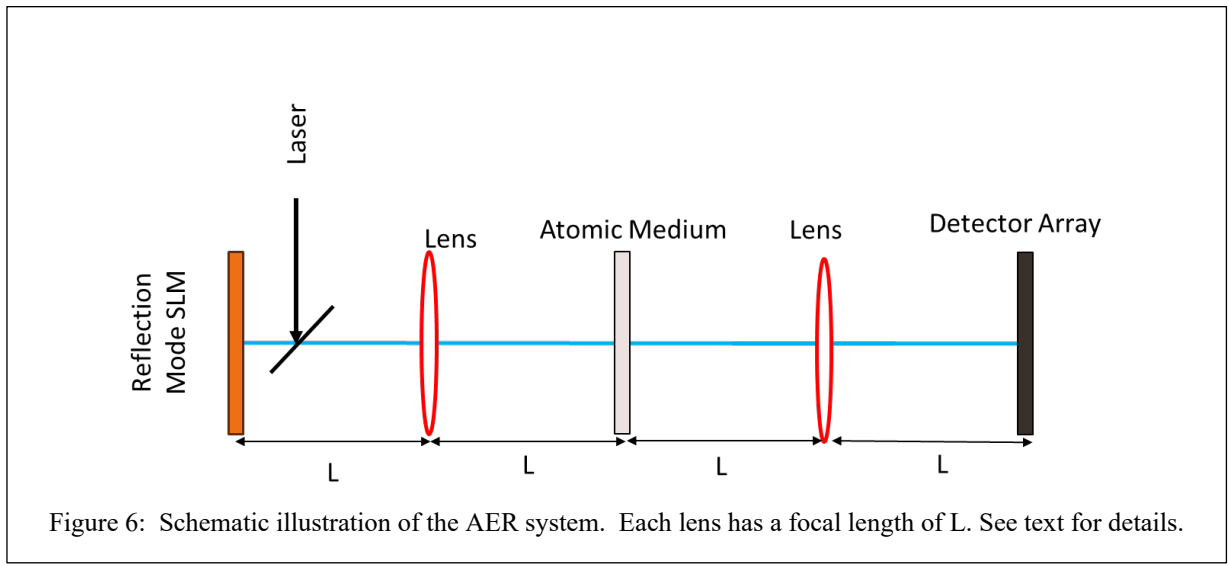

Figure 6: Schematic illustration of the AER system. Each lens has a focal length of L. See text for details.

### 3.1 Applying phase corrections for each pixel from frame to next frame

For each signal, we denote by $W(\vec{\rho})$ the spatial variation of the signal at the plane of the SLM, and by $E(\vec{\rho})$ the spatial variation of the signal at the plane of the atomic medium. Under paraxial approximation and the Fresnel diffraction formula, we get the following relationship between these signal [26]:

$$E(\vec{\rho}) = \frac{\exp(2jkL)}{j\lambda L}\tilde{W}(\vec{f})\Big|_{\vec{f}=\vec{\rho}/\lambda L}, \quad (6)$$

where L is the focus length of the lens, λ is wavelength of light, and $\tilde{W}(\vec{f})$ is the two-dimensional FT of $W(\vec{\rho})$. Thus, apart from the constant phase factor (-jexp (j2kL)) and the scaling factor $(\lambda L)^{-1}$, the field $E(\vec{\rho})$ is the Fourier transform of the input field $W(\vec{\rho})$, as noted earlier. This relationship between the signal at the plane of the atomic medium and the input plane at the SLM must be taken into account carefully when modeling the behavior of the AER, as described next.

Consider first the case where the atomic medium is modeled as an ideal two-level system. In our simulation, we compute the response of the atoms on a pixel-by-pixel form, with the area each pixel chosen to match the spatial resolution of the detector array. As can be seen from Eqn. (6), both the amplitude and the phase of the field will be distinct for each pixel. After the Rotating Wave Approximation (RWA) is applied, the Hamiltonian for each pixel retains time dependence as well as phase dependence. The time dependence can be eliminated by carrying out the Rotating Wave Transformation (RWT). To eliminate the phase dependence and render the Hamiltonian real, we carry out the so-called Q-transformation [27] as follows:

$$\tilde{H} = QHQ^{-1}; \quad |\tilde{\psi}\rangle = Q|\psi\rangle; \quad i\hbar\frac{\partial|\tilde{\psi}\rangle}{\partial t} = \tilde{H}|\tilde{\psi}\rangle; \quad Q = \begin{bmatrix} \exp(-i\phi(x,y)) & 0 \\ 0 & 1 \end{bmatrix} \quad (7)$$

Here, $|\psi\rangle$ is the quantum state and $H$ is the Hamiltonian after carrying out the RWT. The phase factor, $\phi(x,y)$ is potentially distinct for each pixel, denoted by the coordinates $(x,y)$, in the plane of the atomic medium. This factor can change from frame to frame in the video streams representing the query signal and the reference signal.

Consider a situation where the phase at a given pixel has the value of $\phi_A(x,y)$ for one frame (A), and $\phi_B(x,y)$ for the next frame (B). After the interaction with frame A is computed in the atomic medium, it is necessary to account for this phase difference before carrying out the computation for the interaction with the next frame. Specifically, the quantum state must be modified as follows before applying the real Hamiltonian for frame B:

$$|\tilde{\psi}\rangle_B = Q_B Q_A^{-1} |\tilde{\psi}\rangle_A \equiv R_{BA} |\tilde{\psi}\rangle_A; \quad R_{BA} = Q_B Q_A^{-1} = \begin{bmatrix} \exp\{-i[\phi_B(x,y) - \phi_A(x,y)]\} & 0 \\ 0 & 1 \end{bmatrix} \quad (8)$$

When the density matrix equations are used, as necessary for the effective two-level system, the Q-transformation can be expressed as:

$$\tilde{\rho} = Q\rho Q^{-1} \quad (9)$$

The corresponding modification when going from one frame (A) to the next frame (B) (to be applied to the density matrix prior to applying the density matrix equation of motion) can be expressed as follows:

$$\tilde{\rho}_B = Q_B Q_A^{-1} \tilde{\rho}_A Q_A Q_B^{-1} = R_{BA} \tilde{\rho}_A R_{BA}^{-1} \quad (10)$$

For the complete three-level system including effects of decay, we must use the density matrix equations of motion only. In that case, the Q-transformation can be expressed as:

$$\tilde{\rho} = Q\rho Q^{-1}; \quad Q = \begin{bmatrix} \exp(-i\phi^a(x,y)) & 0 & 0 \\ 0 & \exp(-i\phi^b(x,y)) & 0 \\ 0 & 0 & 1 \end{bmatrix} \quad (11)$$

Here, $\phi^a(x,y)$ is the phase of the field at frequency $\omega_a$, and $\phi^b(x,y)$ is the phase of the field at frequency $\omega_b$. The corresponding modification when going from one frame (A) to the next frame (B) can be expressed as follows:

$$\tilde{\rho}_B = Q_B Q_A^{-1} \tilde{\rho}_A Q_A Q_B^{-1} = R_{BA} \tilde{\rho}_A R_{BA}^{-1};$$

$$R_{BA} = Q_B Q_A^{-1} = \begin{bmatrix} \exp\{-i[\phi_B^a(x,y) - \phi_A^a(x,y)]\} & 0 & 0 \\ 0 & \exp\{-i[\phi_B^b(x,y) - \phi_A^b(x,y)]\} & 0 \\ 0 & 0 & 1 \end{bmatrix} \quad (12)$$

For the system we have simulated, the field at frequency frequency $\omega_b$ remains constant, with unchanging phase. As such, $[\phi_B^b(x,y) - \phi_A^b(x,y)] = 0$ in the expression for $R_{BA}$ in this case.

## 3.2 Results

To simulate the AER system, we used an image of the letter "A" as both the reference and the Query signal, each representing an event with a single frame, for simplicity. Figure 7 shows the results for the three level system including the effects of decay. Note that the time axis is reversed, so that $T_1<T_2<T_3$. The correlation signal appears at $t= T_3+T_2-T_1$, as expected. The result shown in the top right is found using the ideal two-level system model, and the result shown in the bottom right is found using the analytical model developed in Ref. [2], showing close agreement. While this agreement is encouraging, it should be noted that the analytical model is only valid for weak Rabi frequencies, as employed here. For the more general situation when many frames would be used for both query and reference signals, it would be impossible to employ the analytical model, and we will have to use a numerical model. We have also carried out the same simulation using the ideal two level model, as well as the effective two-level system model, producing very similar results, except for the amplitudes of the correlation signals, which are the weakest for the three-level system, since the signal is manifested in the amplitude of $|\rho_{13}|$ rather than $|\rho_{12}|$.

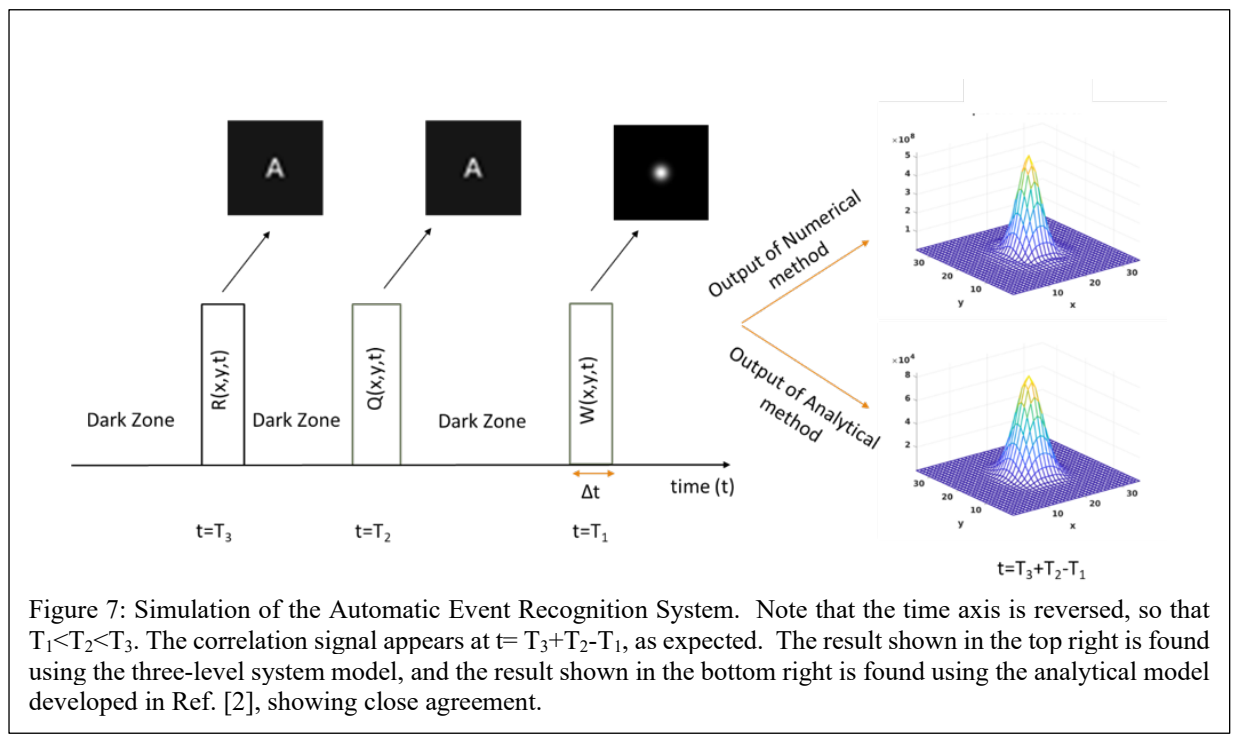

Figure 7: Simulation of the Automatic Event Recognition System. Note that the time axis is reversed, so that $T_1<T_2<T_3$. The correlation signal appears at $t= T_3+T_2-T_1$, as expected. The result shown in the top right is found using the three-level system model, and the result shown in the bottom right is found using the analytical model developed in Ref. [2], showing close agreement.

Figure 8 shows the results when the reference signal is identical to the query signal but with different spatial locations. The first row represents the results when both Query and Reference pulse are centered in spatial domain. The second row shows results when the reference signal is shifted to the right bottom corner in the spatial domain. The third row exhibits the correlation result when the reference signal is shifted to the left-top corner. The numerical results, produced using the three level system model including the effects of decay, are in close agreement with results produced using the analytical model developed in Ref. [2]. We have also carried out the same simulation using the ideal two level model, as well as the effective two-level system model, producing very similar results, except for the amplitudes of the correlation signals, which again are the weakest for the three-level system, since the signal is manifested in the amplitude of $|\rho_{13}|$ rather than $|\rho_{12}|$. Similar simulation will be carried out in the near future using a more realistic situation where both the query signal and the reference signal contain many frames.

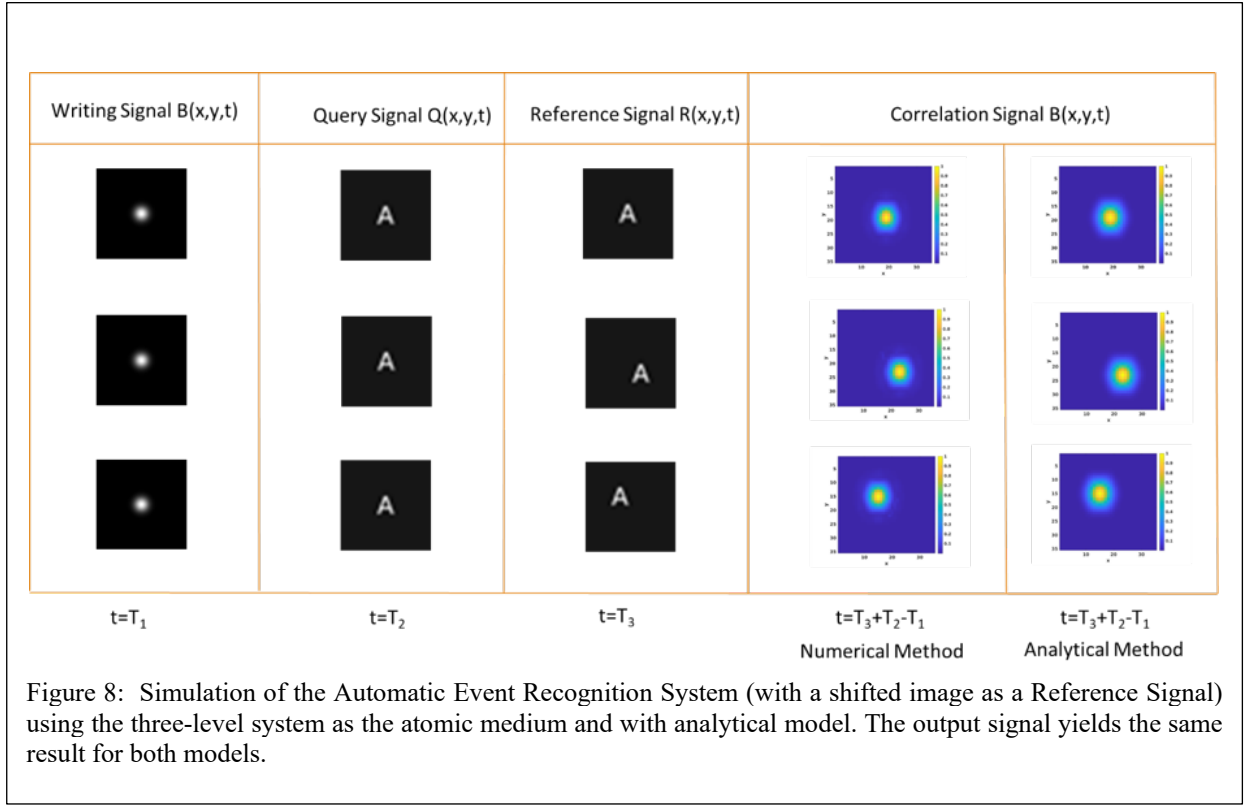

Figure 8: Simulation of the Automatic Event Recognition System (with a shifted image as a Reference Signal) using the three-level system as the atomic medium and with analytical model. The output signal yields the same result for both models.

## 4. Conclusion

We have presented results of simulating the performance of an automatic event recognition system employing an atomic medium of a three-level system, taking into account the effect of decay. Such a system transfers optical coherence to long-lived electro-nuclear spin coherence and vice-versa, thus making it possible to recognize events with long duration, while still using signals carried by optical images. We show that an effective two-level model, generated via adiabatic elimination of the excited state, yields good agreement with the full three-level model. We also show how the temporally and spatially varying phases of the fields impinging on each pixel of the atomic medium are to be taken into account in such models.

## Acknowledgements

The work reported here was supported by the Air Force Office of Scientific Research under Grant Agreements No. FA9550-18-01-0359 and FA9550-23-1-0617.

[27] In practice, the RWT and the phase transformation can be carried out simultaneously. Here, for clarity, we separate these processes, and define the Q transformation as the one used solely for eliminating the phase dependence.